\documentclass{aa}
\usepackage{graphicx}
\begin{document}

\title{Hot gas in Mach cones around Virgo Cluster spiral galaxies$^*$}

\author{
M.~We\.zgowiec\inst{1}
 \and B.~Vollmer\inst{2}
 \and M.~Ehle\inst{3}
 \and R.-J.~Dettmar\inst{1}
 \and D.J.~Bomans\inst{1}
 \and K.T.~Chy\.zy\inst{4}
 \and M.~Urbanik\inst{4}
 \and M.~Soida\inst{4}}
\institute{Ruhr-Universit\"at Bochum, Universit\"atsstrasse 150, 44780 Bochum, Germany
\and CDS, Observatoire astronomique de Strasbourg, UMR7550,
11 rue de l'universit\'e, 67000 Strasbourg, France
\and ESAC, XMM-Newton Science Operations Centre
P.O. Box 78, 28691 Villanueva de la Ca\~nada, Madrid, Spain
\and Obserwatorium Astronomiczne Uniwersytetu
Jagiello\'nskiego, ul. Orla 171, 30-244 Krak\'ow, Poland}

\offprints{M. We\.zgowiec}
\mail{mawez@astro.rub.de\\
$^*$Based on observations obtained with XMM-Newton, an ESA science mission with instruments and contributions
directly funded by ESA Member States and NASA}
\date{Received date/ Accepted date}

\titlerunning{Hot gas in Mach cones around Virgo Cluster spiral galaxies}
\authorrunning{We\.zgowiec et al.}

\abstract
%context
{The detailed comparison between observations and simulations of
ram pressure stripped spiral galaxies in the Virgo cluster has led to
a three dimensional view of the galaxy orbits within the hot
intracluster medium. The 3D velocities and Mach numbers derived
from simulations can be used to derive simple Mach cone geometries for Virgo spiral galaxies.}
%aims
{We search for indications of hot gas within Mach cones in X-ray observations of selected Virgo Cluster spiral galaxies (NGC~4569, NGC~4388, and NGC~4501).}
%methods
{Low-resolution maps of diffuse extended emission and X-ray spectra from XMM-Newton observations are presented. 
Gas densities and temperatures are derived from the X-ray spectra.}
%results
{We find extraplanar diffuse X-ray emission in all galaxies. 
Based on the 3D velocity vectors from dynamical modelling a simple Mach cone is
fitted to the triangular shape of NGC~4569's diffuse X-ray emission. 
Assuming that all extraplanar diffuse X-ray emission has to be located inside the Mach cone, we also
fit Mach cones to NGC~4388's and NGC~4501's extraplanar X-ray emission. For NGC~4569 it is hard to reconcile the derived
Mach cone opening angle with a Mach number based on the sound speed alone. Instead, a Mach number
involving the Alfv\'enic speed seems to be more appropriate, yielding a magnetic field strength of $\sim 3$-$6$~$\mu$G
for a intracluster medium density of $n \sim 10^{-4}$~cm$^{-3}$.
Whereas the temperature of the hot
component of NGC~4569's X-ray halo ($0.5$~keV) is at the high end but typical for a galactic outflow, the temperature
of the hot gas tails of NGC~4388 and NGC~4501 are significantly hotter ($0.7$-$0.9$~keV).}
%conclusions
{In NGC~4569 we find direct evidence for a Mach cone which is filled with hot gas from a galactic superwind.
We suggest that the high gas temperatures in the X-ray tails of NGC~4388 and NGC~4501 are due to the mixing 
of the stripped ISM into the hot intracluster medium of the Virgo cluster.}

\maketitle

\section{Introduction \label{intro}}

There is ample evidence that the cluster environment modifies the gas distribution of spiral galaxies. Especially
in the Virgo cluster, where the gas distribution of spiral galaxies can be observed in detail by current interferometric observations, 
radially truncated gas disk are frequently observed (Cayatte et al. 1990, Chung et al. 2009). The Virgo cluster has
the advantage that it is spiral-rich and dynamically young, i.e. there are strong ongoing environmental interactions.
Chung et al. (2007) found seven spiral galaxies with long H{\sc i} tails in intermediate to low galaxy-density regions 
(0.6-1 Mpc in projection from M87). 
The tails are all pointing roughly away from M87. Assuming radial galaxy orbits this suggests that these tails are 
due to ram pressure stripping. 
Therefore, ram pressure stripping already begins to affect spiral galaxies around the cluster Virial radius. 
The further evolution of a galaxy depends critically on its orbit (see, e.g., Vollmer et al. 2001), 
i.e. a highly eccentric orbit will lead the galaxy at a high velocity into the cluster core, where the intracluster medium is 
densest and ram pressure will be very strong. 
Recently, Vollmer (2009) established a model-based ram pressure stripping time sequence of Virgo spiral galaxies.
Snapshots of three-dimensional models including ram pressure were compared to the observed H{\sc i} gas distributions and
velocity fields. The 3D velocity vector from the dynamical model and the timestep of the best-fit model has
to be consistent with the galaxy's projected position and radial velocity within the Virgo cluster.
Moreover, Vollmer (2009) concluded that the linear orbital
segments derived from the dynamical models together with the intracluster medium density distribution derived
from X-ray observations are consistent with the dynamical simulations of 6 Virgo cluster spirals.
The deduced closest approach of the galaxies to the cluster center (M~87) is between $200$ and $600$~kpc.

Another consequence of ongoing ram pressure can be found in the distribution of polarized radio continuum emission. 
Polarized radio continuum emission is sensitive to gas compression and shear motions. Asymmetric ridges of polarized 
radio continuum emission are frequently observed in Virgo spiral galaxies with disturbed or truncated H{\sc i} disks
(We\.{z}gowiec et al. 2007, Vollmer et al. 2007). MHD modelling based on the velocity fields of the dynamical model
can corroborate a ram pressure stripping scenario and/or discriminate between different interaction scenarios
(NGC~4654, Soida et al. 2006; NGC~4522, Vollmer et al. 2006; NGC~4501, Vollmer et al. 2008).

Despite these clear signs of ram pressure stripping on the atomic gas content, only a handful of gas tails associated to
cluster spiral galaxies have been detected in X-rays: in ACO 2125 (C153 at z = 0.253, Wang et al. 2004),
in ACO 1367 (UGC 6697 at z = 0.022, Sun \& Vikhlinin 2005), in the Coma cluster (NGC~4848, Finoguenov et al. 2004),
and in ACO 3627 (ESO 137-001, Sun et al. 2006). 
During a ram pressure stripping event it is expected that gas is heated to
X-ray temperature by bow shocks, heat conduction, and the mixing of the stripped ISM into the hot intracluster medium 
(e.g., Stevens et al. 1999, Schulz \& Struck 2001, Roediger et al. 2006, Roediger \& Br\"{u}ggen 2008, Tonnesen \& Bryan 2009).
So it is rather surprising and not clear why the X-ray detection rate is so low.
In particular, there is no detection of the leading bow shock of a cluster spiral galaxy in the literature to our 
knowledge (Rasmussen et al. 2006 detected a shock-like feature in NGC~2276 which is part of the
NGC~2300 group of galaxies). Since galaxies move with velocities between $1000$ and $2000$~km\,s$^{-1}$ and the sound speed
in the intracluster medium of the Virgo cluster is $\sim 500$-$700$~km\,s$^{-1}$, a bow shock with an
associated Mach cone is expected.

In this article we use XMM-Newton data to search for X-ray signatures of 3 ram pressure stripped spiral galaxies 
belonging to the ram pressure stripping sequence from Vollmer (2009).
We made deep XMM-Newton spectroscopic observations of NGC~4569 and use archival XMM-Newton data for NGC~4388, and NGC~4501. 
They are described in Sect.~\ref{sec:obsred} and the results are presented in Sect.~\ref{sec:results}.
We find extraplanar diffuse gas in all galaxies. Simple Mach cone geometries are fitted to the
X-ray images and the thermal pressure is compared to ram and cosmic ray pressure in Sect.~\ref{sec:discussion}. 
We give our conclusions in Sect.~\ref{sec:conclusions}.

\section{Observations \label{sec:obsred}}

\begin{table*}
        \caption{Basic astronomical properties of studied galaxies}
        \begin{center}
        \begin{tabular}{ccccccc}
\hline
Name &\vspace{1pt} Morph. & Optical position$^{\rm a}$ &\vspace{1pt} Incl.$^{\rm a}$ &\vspace{1pt} Pos. & Proj. dist. & N$_{\rm H}^{\rm b}$ \\
 & type$^{\rm a}$& \hspace{5pt} $\textstyle\alpha_{2000}$\hspace{30pt}$\textstyle\delta_{2000}$ & [\degr]& ang.$^{\rm a}$[\degr]& to Vir A [\degr] & [$10^{20}$~cm$^{-2}$] \\
\hline
\vspace{5pt}
NGC~4388 & Sb & 12$^{\rm h}$25$^{\rm m}$46\fs8 \hspace{1pt} +12\degr 39\arcmin 44\arcsec & 82 & 91 & 1.3 & 2.58\\
\vspace{5pt}
NGC~4501 & Sb & 12$^{\rm h}$31$^{\rm m}$59\fs3 \hspace{1pt} +14\degr 25\arcmin 14\arcsec & 60 & 138 & 2.0 & 2.62\\
\vspace{5pt}
NGC~4569 & SABa & 12$^{\rm h}$36$^{\rm m}$50\fs1 \hspace{1pt} +13\degr 09\arcmin 46\arcsec & 66 & 23 & 1.7 & 2.82\\
\hline
\end{tabular}
\label{objects}
\end{center}
$^{\rm a}$ taken from HYPERLEDA database -- http://leda.univ-lyon1.fr -- see Paturel et al.~(\cite{leda}).\\
$^{\rm c}$ weighted average value after LAB Survey of Galactic \ion{H}{i}, see Kalberla et al.~(\cite{lab}).\\
\end{table*}

\begin{table*}
\caption{Parameters of X-ray observations of studied galaxies}
\begin{center}
\begin{tabular}{llllllll}
\hline
Name & Obs. ID & Obs. date & Exp. time$^{\rm a}$ & pn filter & pn obs. & MOS filter & MOS obs. \\
   & & & & & ($^{\rm b}$) & & ($^{\rm b}$)  \\
\hline
\vspace{5pt}
NGC~4388 & 0110930701 & 2002-12-12 & 11.9 (6.9) & Thin & EF & Medium & FF \\
\vspace{5pt}
NGC~4501 & 0112550801 & 2001-12-04 & 14 (2.9) & Medium & EF & Thin & FF \\
\vspace{5pt}
NGC~4569 & 0200650101 & 2004-12-13/14 & 66 (49) & Thin & EF & Medium & FF \\
\hline
\end{tabular}
\label{xobs}
\end{center}
$^{\rm a}$ Total time in ksec with clean time for pn camera in brackets.\\
$^{\rm b}$ Observing mode: FF - Full Frame, EF - Extended Full Frame.\\
\end{table*}

We made deep XMM-Newton observations of NGC~4569 (proposal ID: 2006501, PI: M.~Ehle). In addition,
archival XMM-Newton observation of NGC~4388 and NGC~4501 are used.
The main astronomical parameters of the studied galaxies are presented in Table~\ref{objects} and the properties of their X-ray observations are presented in Table~\ref{xobs}. 
The data was processed using the SAS 9.0 package (Gabriel et al.~\cite{sas})
with standard reduction procedures. 
Following the routine of tasks $epchain$ and $emchain$
calibrated event lists for the pn camera (Str\"uder et al.~\cite{strueder}) and  
for the two MOS cameras (Turner et al.~\cite{turner}) were obtained for each galaxy. 
Next, the event lists were carefully filtered for bad CCD
pixels and periods of intense radiation of high energy background.
The filtered event lists were used to produce images, background images, exposure maps (without and
with vignetting correction), masked for an acceptable detector area using the images
script\footnote{http://xmm.esac.esa.int/external/xmm\_science/\\gallery/utils/images.shtml}.
All images and maps were produced in the band of 0.2 - 1~keV in order to
be sensitive for the softest X-ray emission due to hot gas.
Next, the final images were combined using the data from all cameras. The resulting images were adaptively smoothed with a Gaussian beam of 10$\arcsec$
HPBW. To get a better signal to noise ratio and thus better sensitivity for extended structures, which allowed to bring up the large-scale structure,
the images were again smoothed, this time with a Gaussian beam of 30$\arcsec$ or 1$\arcmin$ using the AIPS package. 
The gaps between instrument detectors were not interpolated, but filled by combining scaled MOS and pn maps.

For the spectral analysis, the images were searched for point-like sources using the routine task $edetect\_chain$. Next, after
excluding circular areas around the detected sources from the relevant region area a spectrum was acquired.
Sizes of circular areas were determined by visual inspection of an image of each point source in the pn event list.
Blank sky event lists (see Carter \& Read~\cite{carter}) were used for creating background spectrum corresponding to a proper region.
For each spectrum response matrices and effective area files were produced.
For the latter, detector maps needed for extended emission analysis
were also created. Finally, the spectra were then fitted using XSPEC~11 (Arnaud 1996).
The thermal plasma MEKAL model is based on calculations by Mewe and Kaastra~(Mewe et al.~\cite{mewe},
Kaastra~\cite{kaastra}).

\section{Results \label{sec:results}}

In the following we present the X-ray distribution in the band of 0.2 - 1~keV and X-ray spectra of
NGC~4569, NGC~4388, and NGC~4501 from regions presented in Appendix A.
The unabsorbed X-ray fluxes, X-ray luminosities, gas densities and temperatures are presented in Table~\ref{tab:tempdens}.
The volumes of emitting regions were estimated in the following way (see Appendix A):
(i) the galactic disk regions are approximated by cylinders with radii set by the extent of X-ray emission and
a height of $1$~kpc, (ii) the halo emission of NGC~4569 is assumed to be spherical with a radius of $11.4$~kpc,
(iii) the tail region of NGC~4388 is approximated by a cylinder with a length of $50.8$~kpc and a radius of $9.3$~kpc,
and (iv) the tail region of NGC~4501 is approximated by a cylinder with a length of $30$~kpc and a radius of $14$~kpc.
The model types and reduced $\chi^2$ are shown in Table~\ref{tab:models}.
The volumes are uncertain by roughly a factor of 2. The typical uncertainty of the density is $30-50$\,\%,
that of the masses is $\sim 0.2$~dex.
The volumes and gas masses of the different X-ray emitting components are presented in Table~\ref{tab:tempdens1}.

\begin{table*}
       \caption{X-ray fluxes, derived densities, and temperatures from the X-ray spectra.}
       \begin{tabular}{cccccccc}
\hline
Galaxy  & Region & hot gas flux$^{{\rm a}}$                   & L$_{\rm X}$              & $(n_{\rm cold}\ \eta^{0.5})^{\rm b}$      & $kT_{\rm cold}$          & $(n_{\rm hot}\ \eta^{0.5})^{\rm b}$            & $kT_{\rm hot}$ 	  \\
        &        &[10$^{-14}$ erg\,cm$^{-2}$s$^{-1}$]      & [10$^{39}$erg\,s$^{-1}$] & [$10^{-3}{\rm cm}^{-3}$]    & [keV]                    & [$10^{-3}{\rm cm}^{-3}$] & [keV]          \\
\hline
NGC     & disk   &  2.26$^{+1.48}_{-1.35}$                                        &     0.780                &  1.6$^{+1.2}_{-0.8}$       & 0.14$^{+0.10}_{-0.04}$   &  2.0$^{+0.5}_{-0.6}$   & 0.46$\pm$0.08  \\
4569    & halo   & 11.39$^{+5.28}_{-4.39}$                                        &     3.936                &  0.7$^{+0.3}_{-0.2}$  & 0.17$^{+0.02}_{-0.03}$   &  0.7$^{+0.2}_{-0.2}$    & 0.49$\pm$0.05  \\
\hline
NGC     & nucleus & 16.84$^{+6.40}_{-8.92}$                                         &     5.819                & 27.8$^{+13.7}_{-10.6}$  & 0.14$^{+0.07}_{-0.03}$           & 18.8$^{+5.2}_{-6.2}$    & 0.66$^{+0.10}_{-0.04}$ \\
4388    & disk   & 10.44$^{+9.77}_{-10.44}$                                      &     3.608                &  4.6$^{+1.3}_{-1.3}$                & 0.21$\pm$0.03                    &  3.7$^{+1.0}_{-1.0}$  & 0.79$\pm$0.11  \\
        & tail   & 21.51$^{+5.92}_{-8.82}$                                        &     7.433                &    0.7$^{+0.3}_{-0.3}$   &   0.87$^{+0.13}_{-0.12}$                           &      0.9$^{+2.6}_{-0.9}$ & $2.26^{+2.16}_{-0.85}$ \\
\hline
NGC     & disk   & 16.44$^{+8.31}_{-6.03}$                                       &     5.681                &  --                & --                              &  3.4$^{+1.0}_{-1.0}$             & 0.38$^{+0.12}_{-0.07}$ \\
4501    & tail   & 14.31$^{+4.25}_{-8.87}$                                     &     4.945                &  --                & --                              &  0.6$^{+0.2}_{-0.2}$             & 0.67$^{+0.11}_{-0.09}$ \\
\hline
\end{tabular}
             \\
$^{\rm a}$ Total unabsorbed flux from the thermal emission between 0.3 and 12~keV.\\
$^{\rm b}$ $\eta$ is the volume filling factor.\\
\label{tab:tempdens}
\end{table*}

\begin{table}
       \caption{Estimated volumes and gas masses.}
       \begin{tabular}{ccccc}
\hline
Galaxy  & Region & volume & (M$_{\rm gas}^{\rm cold} \eta^{0.5}$)$^{\rm a}$ & (M$_{\rm gas}^{\rm hot} \eta^{0.5}$)$^{\rm a}$ \\
 & & [pc$^{3}$] & [M$_{\odot}$] & [M$_{\odot}$] \\
\hline
NGC    & disk    & 1.77e+11 & 6.6$^{+5.6}_{-4.6}$e+06 & 7.8$^{+4.5}_{-4.6}$e+06 \\
4569 & halo  & 6.14e+12 & 9.8$^{+6.0}_{-5.7}$e+07 & 1.0$^{+0.6}_{-0.6}$e+08 \\
\hline
NGC & nucleus & 1.10e+10 & 6.9$^{+4.8}_{-4.3}$e+06 & 4.7$^{+2.7}_{-2.8}$e+06 \\
4388 & disk & 4.52e+11 & 4.6$^{+2.7}_{-2.7}$e+07 & 3.7$^{+2.1}_{-2.1}$e+07 \\
 & tail & 1.38e+13 & 2.1$^{+1.4}_{-1.4}$e+08 & 2.8$^{+8.1}_{-3.1}$e+08 \\
\hline
NGC & disk & 7.10e+11 & - & 5.4$^{+3.2}_{-3.1}$e+07 \\
4501 & tail & 1.91e+13 & - & 2.5$^{+1.4}_{-1.4}$e+08 \\
\hline
\end{tabular}
$^{\rm a}$ $\eta$ is the volume filling factor.\\
\label{tab:tempdens1}
\end{table}

\begin{table*}
    \caption{Model type and reduced $\chi_{\rm red}^2$.}
       \begin{tabular}{lcc}
\hline
Galaxy  & model type & $\chi_{\rm red}^2$ \\
\hline
NGC~4569 disk & wabs(mekal+mekal+powerlaw) & 0.96 \\
NGC~4569 halo & wabs(mekal+mekal+powerlaw) & 1.23 \\
NGC~4388 nucleus & wabs(mekal+mekal+powerlaw+wabs(powerlaw+gauss)) & 1.06 \\
NGC~4388 disk+nucleus & wabs(mekal+mekal+powerlaw+wabs(powerlaw+gauss)) & 1.07 \\
NGC~4388 tail & wabs(mekal+mekal+powerlaw) & 1.11 \\
NGC~4501 disk & wabs(mekal+powerlaw) & 1.39 \\
NGC~4501 tail & wabs(mekal+powerlaw) & 0.46 \\
\hline
\end{tabular}
\label{tab:models}
\end{table*}

\subsection{NGC~4569 \label{4569}}

\begin{figure}[ht]
\resizebox{\hsize}{!}{\includegraphics{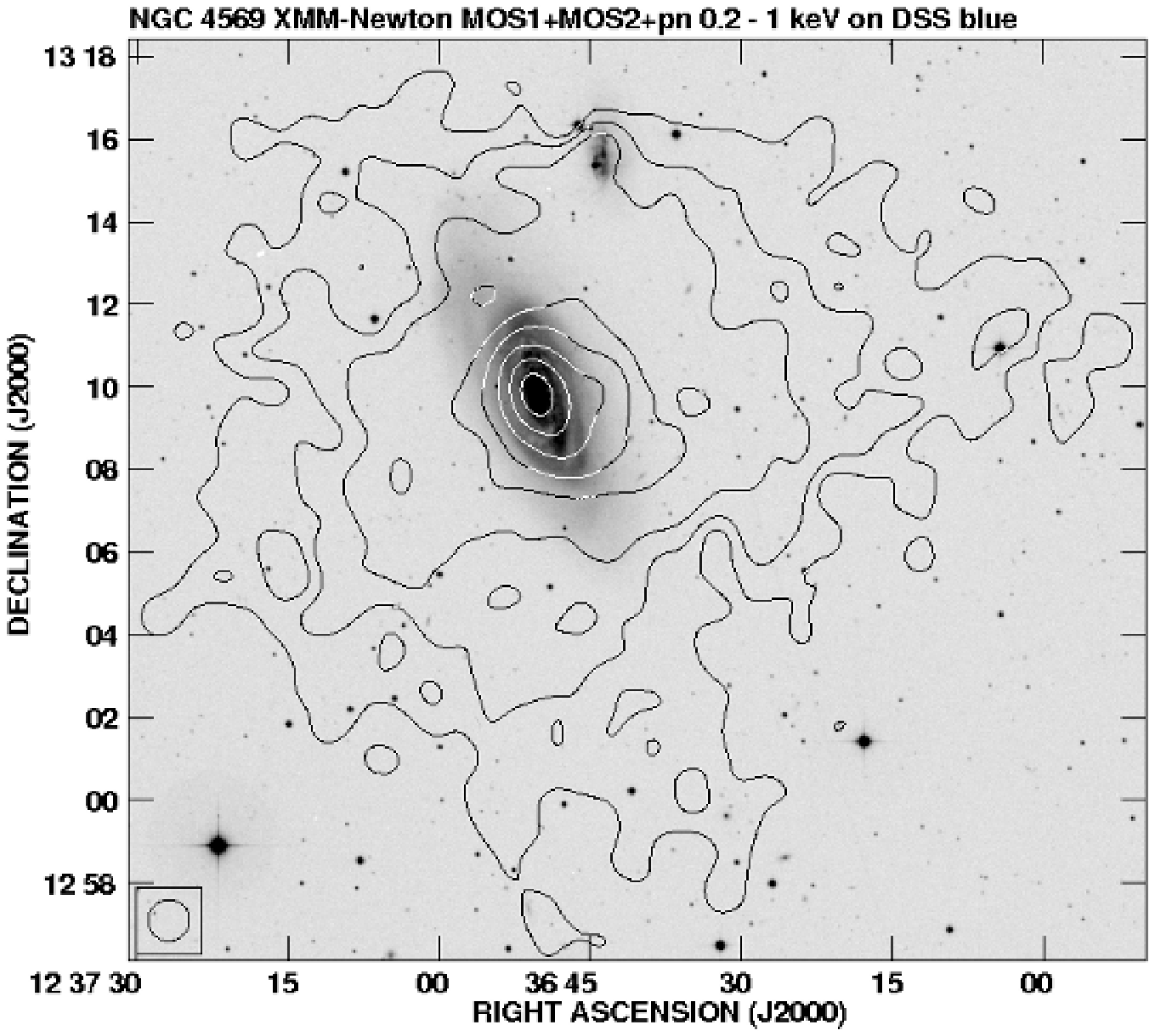}}
\resizebox{\hsize}{!}{\includegraphics{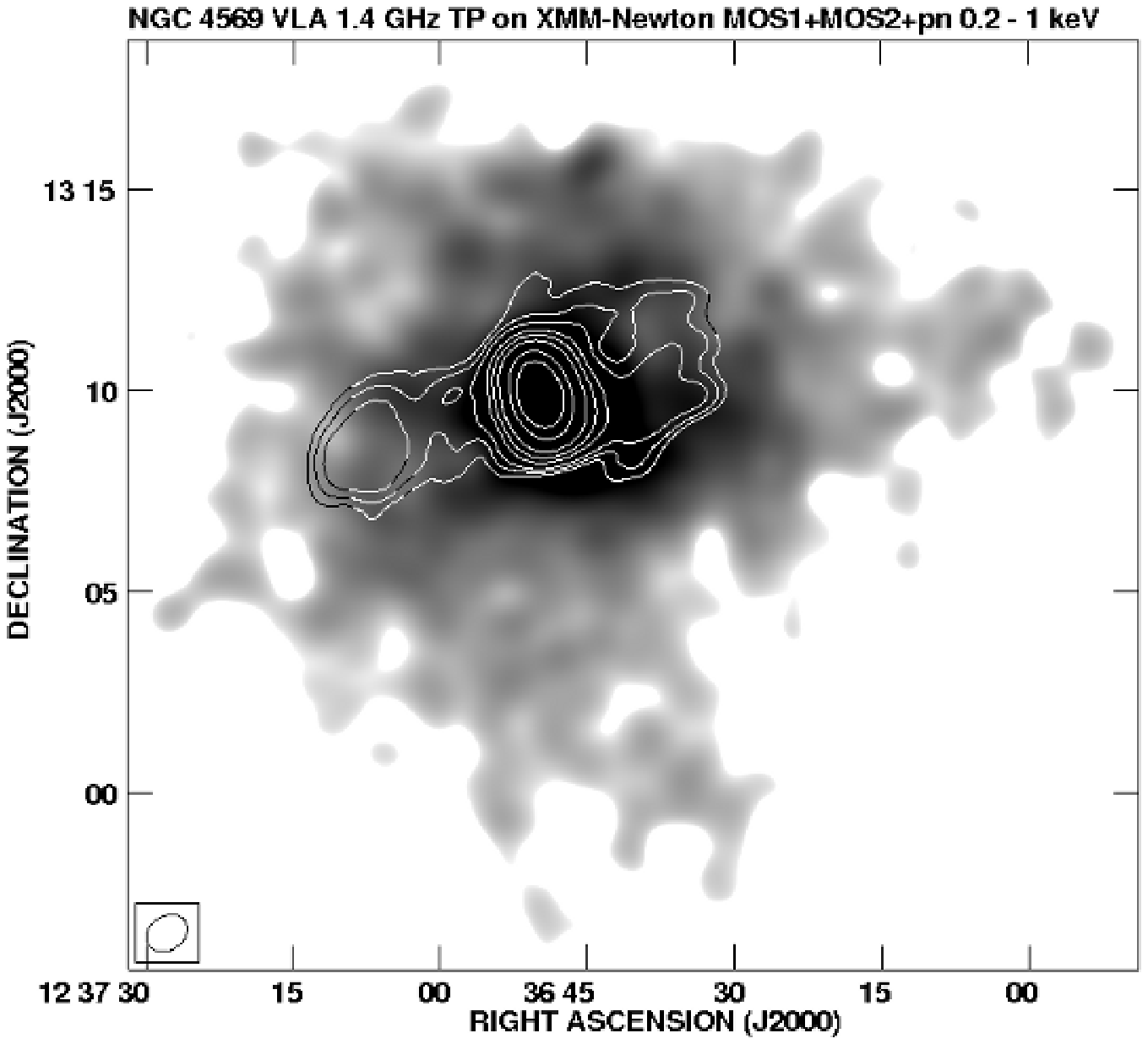}}
\caption{X-ray emission (0.2-1~keV) of NGC~4569 together with an optical B band (upper panel) and the
radio total power map at 1.4~GHz from Chy\.zy et al.~(\cite{chyzy4569}; lower panel).
The resolution of both maps is 1$\arcmin$. The beam size is shown in the bottom left corner of the figure.
The first greyscale in the lower panel corresponds to 3 times the rms noise level.
The radio contour levels are 3, 5, 8, 16, 25, 40, 80, 120$\times 0.24$~mJy.}
\label{fig:n4569x}
\end{figure}

The diffuse X-ray emission associated to NGC~4569 is extended over an area of approximately
$20' \times 20'$ ($100 \times 100$~kpc; Fig.~\ref{fig:n4569x}).
%of $2.5' \times 2.5'$ ($12 \times 12$~kpc)
The overall shape of the X-ray emission is that of a triangle with the tip pointing
to the northeast. The galactic disk has a 5 times lower X-ray flux than the giant diffuse halo
which encompasses also a dwarf satellite galaxy IC~3583.
The large radio lobes observed at 20~cm (Chy\.{z}y et al. 2006) are located entirely within the diffuse
X-ray halo. There is enhanced X-ray emission associated to the southern half of the western radio lobe
(at RA=12h36m30s DEC=$13^{\circ} 09'$).

\begin{figure}[ht]
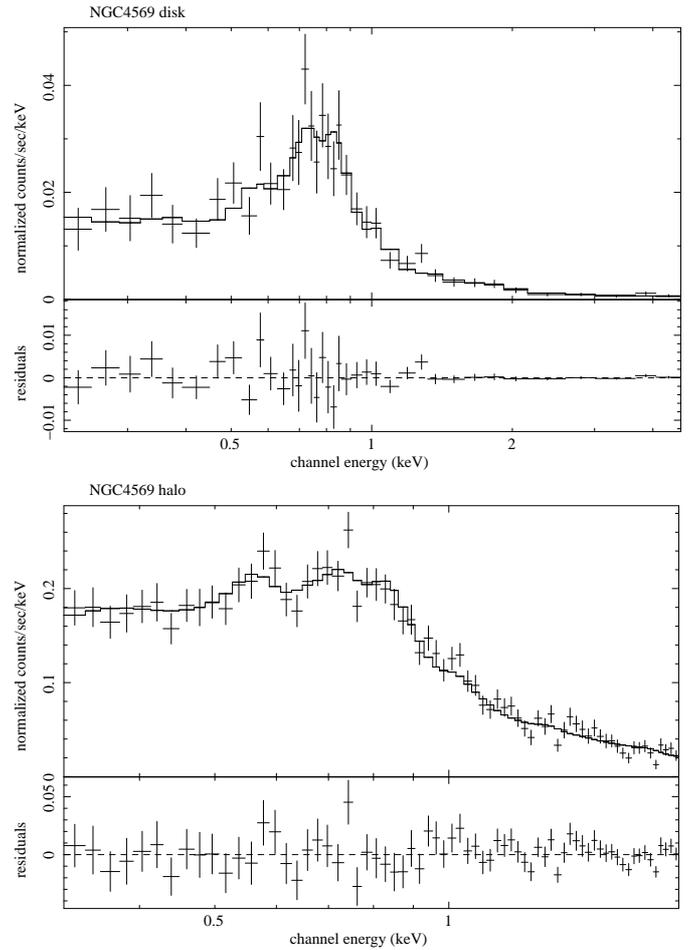

\resizebox{\hsize}{!}{\includegraphics[angle=-90]{16344_f2a.ps}}
\resizebox{\hsize}{!}{\includegraphics[angle=-90]{16344_f2b.ps}}
\caption{X-ray spectrum of NGC~4569's disk (upper panel) and inner halo excluding emission from the galactic disk
(lower panel) .}
\label{fig:n4569spec}
\end{figure}

X-ray spectra (Fig.~\ref{fig:n4569spec}) have been taken from the disk and the inner halo ($2.5' \times 2.5'$ or $12 \times 12$~kpc)
excluding emission from the galactic disk.
They are fitted by a two temperature thermal plasma MEKAL model with a hot and a cold component. 
An additional powerlaw component is added to account for undetected point sources. Both, disk and halo
show a cold ($\sim 0.15$~keV) and a hot component ($\sim 0.5$~keV) and densities of $n \sim 10^{-3}\,\eta^{-0.5}$~cm$^{-3}$
(Table~\ref{tab:tempdens}), where $\eta$ is the volume filling factor.

\subsection{NGC~4388 \label{4388}}

\begin{figure}[ht]
\resizebox{\hsize}{!}{\includegraphics{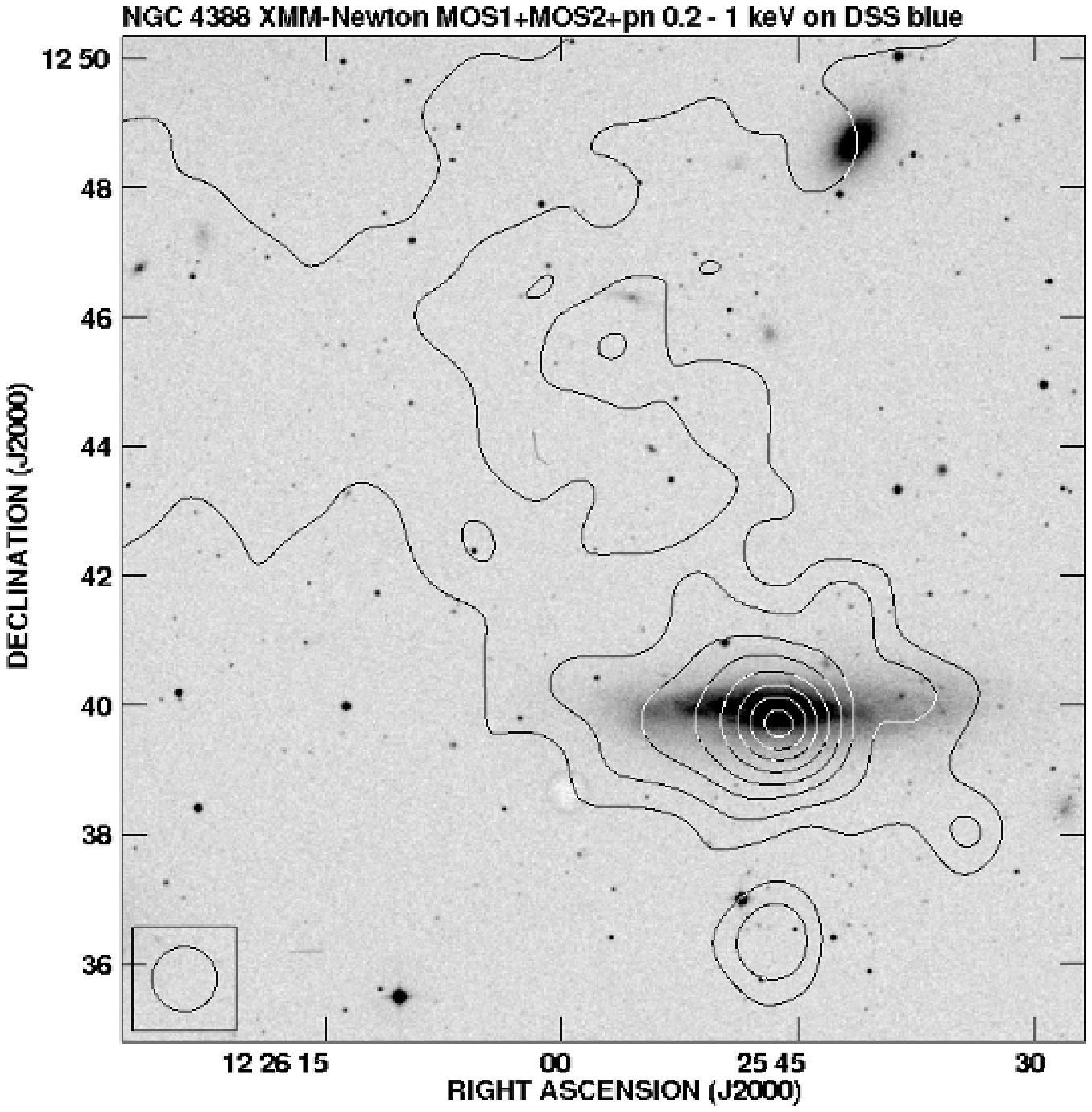}}
\resizebox{\hsize}{!}{\includegraphics{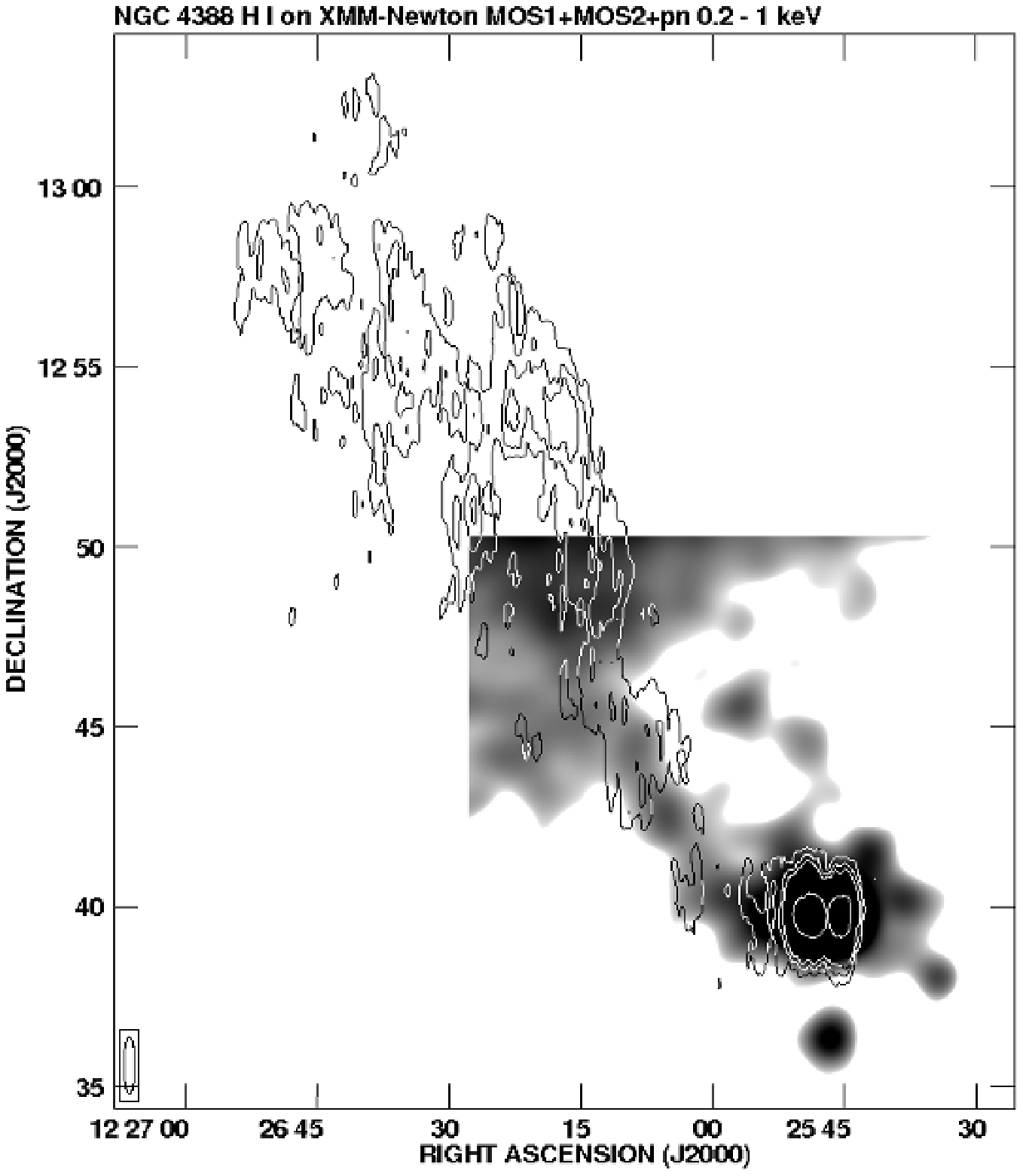}}
\caption{X-ray emission (0.2-1~keV) of NGC~4388 together with an optical B band image (upper panel)
and the H{\sc i} map from Oosterloo \& van Gorkom~(2005; lower panel).
The X-ray map was cut at the position, where contribution from the M~86 halo starts to dominate.
The first greyscale corresponds to 7 times the rms noise level.}
\label{fig:n4388x}
\end{figure}

The galactic disk of NGC~4388 is detected in X-rays. It is more extended to the north than to the south.
In addition, an extended tail is detected towards the northeast of the galactic disk. This X-ray tail is associated with the
H$\alpha$ plume reported by Yoshida et al.~(\cite{yoshida2002}) and the beginning of the huge H{\sc i} tail
discovered by Oosterloo \& van Gorkom (2005) (Fig.~\ref{fig:n4388x}). 
At distances $> 5'$ from the galactic disk the X-ray emission starts to be dominated by the bright halo of M~86.

\begin{figure}[ht]
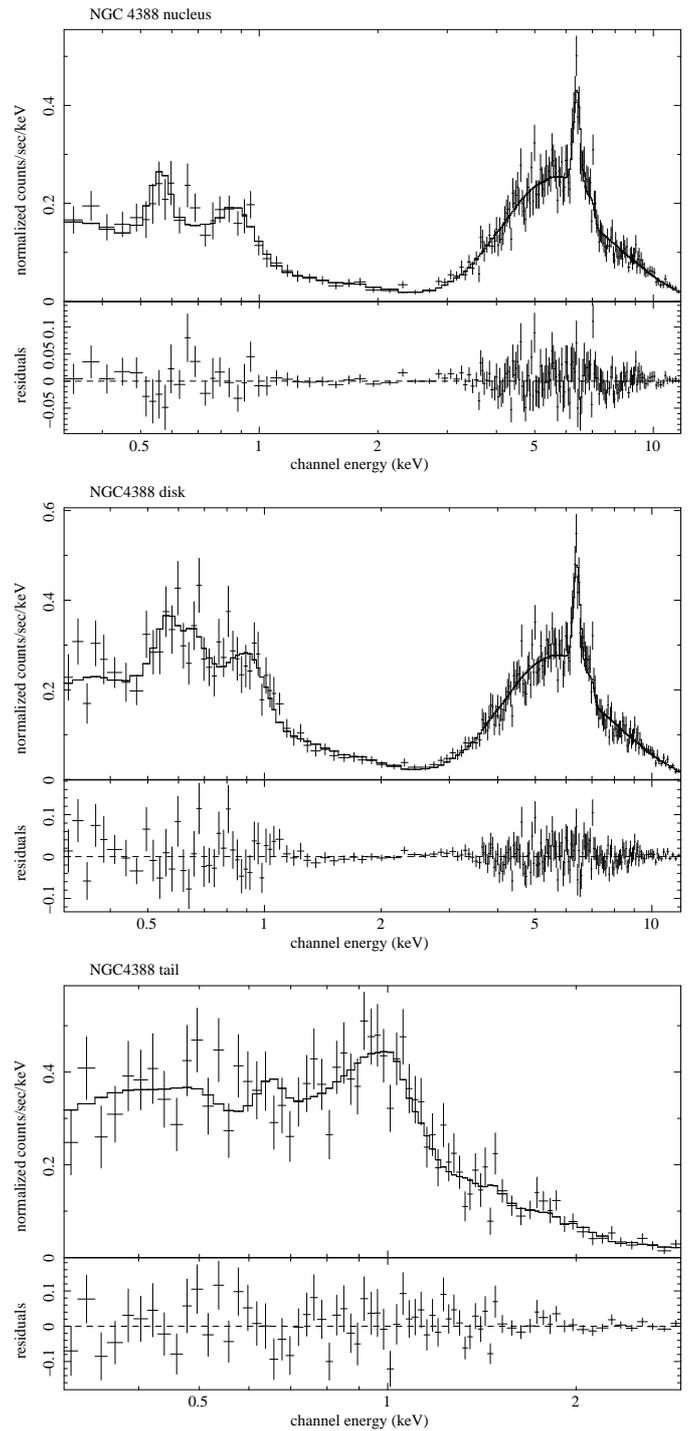

\begin{center}
\resizebox{\hsize}{!}{\includegraphics[angle=-90]{16344_f4a.ps}}
\resizebox{\hsize}{!}{\includegraphics[angle=-90]{16344_f4b.ps}}
\resizebox{\hsize}{!}{\includegraphics[angle=-90]{16344_f4c.ps}}
\end{center}
\caption{The X-ray spectra of NGC~4388's nucleus (upper panel), disk and nucleus (middle panel), and tail (lower panel).}
\label{fig:n4388spec}
\end{figure}

We performed a spectral analysis of the nucleus, the disk and nucleus, and the part of the tail close to the galactic disk. 
The spectra (Fig.~\ref{fig:n4388spec}) are fitted with two single temperature thermal plasma MEKAL models with an additional 
powerlaw component to account for undetected point sources. In the fit of the disk and the nucleus, 
a contribution from the AGN was included by adding to the model an absorbed powerlaw component and the 6.4 keV iron line 
(see Table~\ref{tab:models}).
The temperatures and densities are given in Table~\ref{tab:tempdens}.
The nucleus has a gas density of $\sim 30\,\eta^{-0.5}$~cm$^{-3}$. The cold and hot components of the disk have
temperatures of $\sim 0.2$~keV and $\sim 0.8$~keV, respectively. 
The exceptionally high temperature of the hot component in the disk is probably due to the
nuclear outflow (Veilleux et al. 1999, Yoshida et al. 2004).
In the tail the temperature of the cold component is comparable to the hot component of the disk.
The hot component corresponds to the intracluster medium.
About 60\,\% of the emission comes from the hotter spectral component.

\subsection{NGC~4501 \label{4501}}

\begin{figure}[ht]
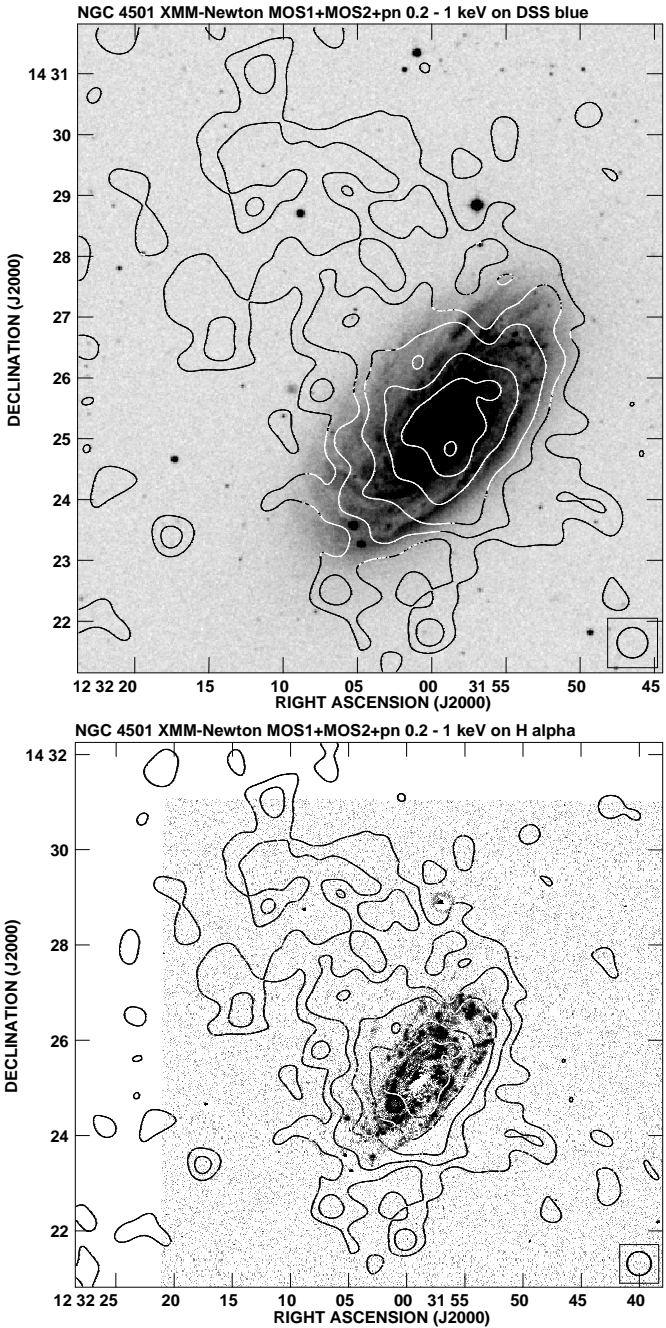

\resizebox{\hsize}{!}{\includegraphics{16344_f5a.ps}}
\resizebox{\hsize}{!}{\includegraphics{16344_f5b.ps}}
\caption{X-ray emission (0.2-1~keV) of NGC~4501 overlaid onto an optical B band image (upper panel) and
the H$\alpha$ image from Goldmine (lower panel).
The map resolution is 30$\arcsec$. The beam size is shown in the bottom right corner of the figure.
The X-ray contours are 3, 5, 8, 16, 25 times the rms noise level.}
\label{fig:n4501x}
\end{figure}

NGC\,4501 shows a moderately extended hot gas halo (Fig.~\ref{fig:n4501x}).
The northeastern side is more extended than the southwestern side.
X-ray emission is found up to $\sim 3'$ ($\sim 15$~kpc) from the galactic disk north from the galaxy center.
This X-ray plume is close to the H$\alpha$-bright spiral arm at the outer northwestern disk
which has no counterpart in the southeast.

\begin{figure}[ht]
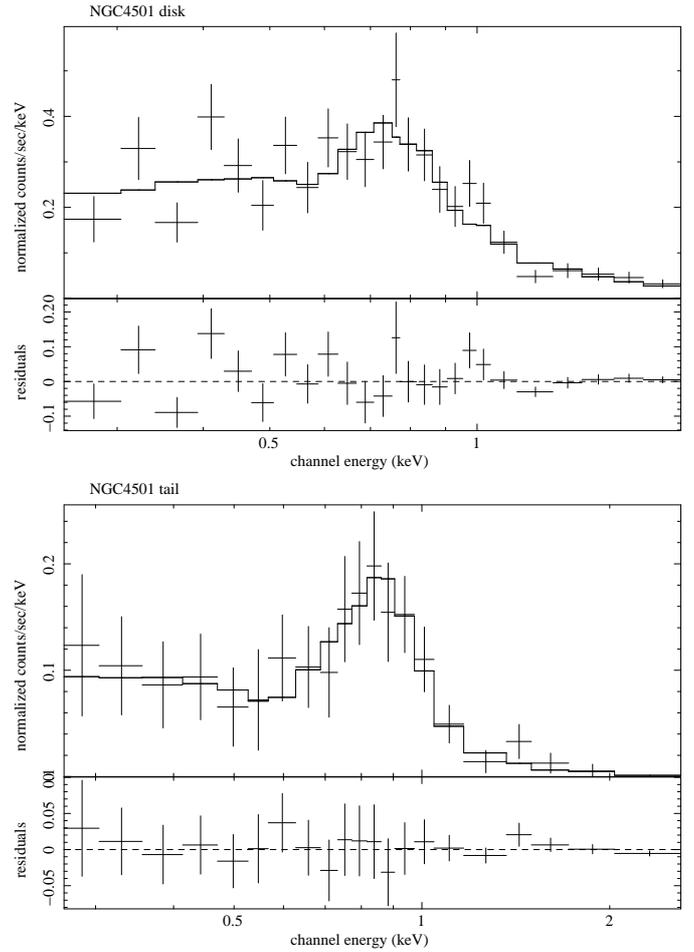

\begin{center}
\resizebox{\hsize}{!}{\includegraphics[angle=-90]{16344_f6a.ps}}
\resizebox{\hsize}{!}{\includegraphics[angle=-90]{16344_f6b.ps}}
\end{center}
\caption{The X-ray spectra of NGC~4501's disk (upper panel) and tail (lower panel).}
\label{fig:n4501spec}
\end{figure}

Due to the limited number of counts, the X-ray spectra of the disk and the tail (Fig.~\ref{fig:n4501spec}) 
has been fitted by a single hot component of $\sim 0.4$~keV and $\sim 0.7$~keV, respectively (Table~\ref{tab:tempdens}).

\section{Discussion \label{sec:discussion}}

To put the observed X-ray emission and spectra of the 3 galaxies into context, we first give a brief summary
on the evolutionary stage of each galaxy.

\subsection{NGC~4569}

The spiral galaxy NGC~4569 is a very peculiar member of the Virgo cluster. Its very large
size ($D_{25}=9.5'$, or 47~kpc) compared to
other cluster members, and negative radial velocity ($v_{\rm r}=-235$~km\,s$^{-1}$ 
with respect to the cluster mean velocity of $\sim 1100$~km\,s$^{-1}$) have lead
to some doubt about its cluster membership, despite its close projected
distance to the cluster center ($d=1.7^{\rm o}=505$~kpc). Van den Bergh (1976) classified this
very red galaxy as anemic due to its low arm--interarm contrast on optical images. 
H{\sc i} line synthesis observations (Warmels 1988; Cayatte et al. 1990) showed that it has lost 
more than 90\% of its H{\sc i}  gas and that its H{\sc i} distribution is heavily truncated.
NGC~4569 is thus an exceptional galaxy, it is one of the largest and most H{\sc i} deficient 
galaxies in the Virgo cluster (Solanes et al. 2001).
 
More recent H{\sc i} observations showed a low surface density arm in the west of the galaxy, 
whose velocity field is distinct from that of the overall disk rotation. 
The observed gas distribution, velocity field, and velocity 
dispersion are consistent with a major
stripping event that took place about 300~Myr ago. This post--stripping scenario can reproduce 
the main observed characteristics of NGC~4569 (Vollmer et al. 2004).
The analysis of the multiwavelength photometry of the galactic disk including GALEX UV observations
(Boselli et al. 2006) confirmed this scenario.
Tsch\"{o}ke et al. (2001) found a diffuse soft X-ray component extended to the west coinciding with a giant H$\alpha$ 
structure. They concluded that the X-ray gas traces a large scale outflow from accumulating supernova explosions 
and stellar winds in the galactic center.

Surprisingly, large radio continuum lobes were found in NGC~4569 by Chy\.{z}y et al. (2006).
These lobes extend up to 24 kpc from the galactic disk and are the first observed in a cluster spiral galaxy.
Since the radio lobes are rather symmetric, they appear to resist ram pressure. 
The most plausible explanation for the origin of the radio lobes is a galactic superwind.
To resist ram pressure, the pressure within the region of the radio lobes has to be
higher than ram pressure.
Assuming the minimum energy condition, the total pressure of cosmic rays and magnetic fields in the lobes
is $p \sim 10^{-12}$~dyn\,cm$^{-2}$. With a mean outflow velocity of $v_{\rm wind}=700$~km\,s$^{-1}$ the mean density is
$n \sim 3 \times 10^{-4}$~cm$^{-3}$ (Chy\.{z}y et al. 2006). 
A sufficiently strong superwind-type flow can easily overcome the gravitational potential of the galaxy and resist ram pressure.
The fact that the radio lobes are entirely engulfed by the diffuse X-ray halo of NGC~4569 lends support 
to this scenario.

The thermal pressure from the halo X-ray spectrum (Fig.~\ref{fig:n4569spec}) can be compared to the pressure
of the cosmic rays and magnetic fields from Chy\.{z}y et al. (2006).
We find a pressure of $p_{\rm th}^{\rm cold} \sim 2 \times 10^{-13}\eta^{-0.5}$~dyn\,cm$^{-2}$ for the cold component
($T=2 \times 10^{6}$~K) and $p_{\rm th}^{\rm hot} \sim 5.5 \times 10^{-13} \eta^{-0.5}$~dyn\,cm$^{-2}$ for the hot component
($T=5.7 \times 10^{6}$~K), where $\eta$ is the volume filling factor. 
The thermal pressure of the hot phase is thus comparable, but a factor of 2 lower than that of the 
cosmic rays and magnetic fields. Thus the superwind might be driven by cosmic rays.

The triangular shape of NGC~4569's diffuse X-ray halo reminds that of a bow shock. From the dynamical model
of Vollmer et al. (2005) and Vollmer (2009) we know the direction of the galaxies motion within the
intracluster medium. With an estimated galaxy velocity of $v_{\rm gal}=1500$~km\,s$^{-1}$ and a sound speed of
$c=550$-$700$~km\,s$^{-1}$ (depending on the adiabatic index) in the intracluster medium the Mach number is 
$M \sim 2.1$-$2.7$ and  the angle of the Mach cone 
(half of the opening angle) is $\alpha=\sin^{-1}(\frac{1}{M})=22$-$28^{\circ}$.
On the other hand, since the very tenuous ($n \sim 10^{-4}$~cm$^{-3}$) intracluster medium is fully ionized and magnetized, 
the sound speed has to be replaced by the magnetosonic velocity $v_{\rm ms}=\sqrt{(c^2+v_{\rm A}^2)}$, where the
Alfv\'{e}n speed is $v_{\rm a}=\frac{B}{\sqrt{4\pi n m_{\rm p}}}$.
Typical magnetic field strengths in galaxy clusters are of the order of a few $\mu$G (Ferrari et al. 2008).
%, it is probable that the gas only becomes collisional 
%because of the magnetic field. The situation might thus be similar to that in the solar wind.
%In  this case the sound speed has to be replaced by the Alfv\'{e}n speed $v_{\rm a}=\frac{B}{\sqrt{4\pi n m_{\rm p}}}$.
%Typical magnetic field strengths in galaxy clusters are of the order of a few $\mu$G (Ferrari et al. 2008). 

\begin{figure}[ht]
\resizebox{\hsize}{!}{\includegraphics{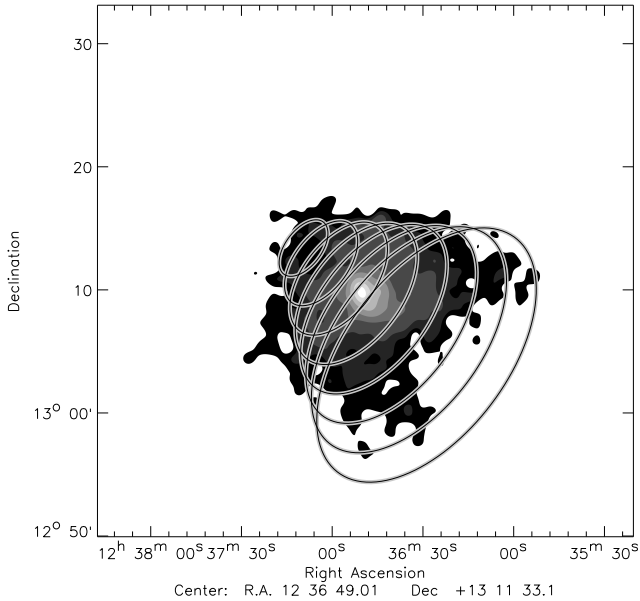}}
\caption{X-ray emission (0.2-1~keV) of NGC~4569 together with the Mach cone
whose direction is based on the dynamical model and whose opening angle is adjusted to the X-ray distribution.}
\label{fig:n4569m}
\end{figure}

We overlaid a Mach cone on the X-ray emission distribution of NGC~4569 in Fig.~\ref{fig:n4569m}.
The direction is based on the dynamical model whereas the opening angle is fitted by eye to the
triangular form of the X-ray distribution.
In this way we find an apparent opening angle of $75^{\circ}$ or an
apparent Mach cone angle of $\alpha=37.5^{\circ}$.
The true Mach cone angle depends on the projection under which the Mach cone is observed, i.e.
the angle $\beta$ between the galaxy's normalized 3D velocity vector and the plane of the sky.
Given the normalized 3D model velocity vector (-0.64,0.53,0.56) (Vollmer 2009), 
this angle is $\beta=34^{\circ}$ and the true Mach cone angle is $30^{\circ}$.
For $\beta=20^{\circ}/40^{\circ}$ the true Mach cone angles are $36^{\circ}$ and $28^{\circ}$, respectively.
NGC~4569's high radial velocity with respect to the Virgo cluster mean favors higher values of $\beta$.
The true Mach cone angle is thus consistent, but might be somewhat higher than our estimates based on the sound speed of the
intracluster medium. We thus suggest that the magnetic field plays a role and that the
magnetosonic velocity is relevant for the determination of the Mach number. 
Based on true Mach cone angles between $30^{\circ}$ and $32^{\circ}$, a sound speed of $550$-$700$~km\,s$^{-1}$, 
a galaxy velocity of $1500$~km\,s$^{-1}$, 
and assuming an intracluster density of $n=10^{-4}$~cm$^{-3}$, we derive Alfv\'en speeds between
$300$~km\,s$^{-1}$ and $600$~km\,s${-1}$ and a magnetic field strength of $\sim 1$-$3$~$\mu$G 
at the location of NGC~4569. The derived range of Alfv\'en speeds is thus comparable to that of the sound speed. 
The derived magnetic field strength is consistent with typical magnetic field strengths in galaxy clusters 
(Ferrari et al. 2008).

\subsection{NGC~4388}

NGC~4388's projected location within the Virgo cluster is close to M~87 ($1.3^{\circ}=385$~kpc).
Yoshida et al.~(\cite{yoshida}) discovered an H$\alpha$ plume which extends up to $35$~kpc northeast
from the galaxy center. Based on a Effelsberg 100m telescope H{\sc i} detection of the H$\alpha$
plume, Vollmer \& Huchtmeier (2003) made dynamical models and concluded that ram pressure stripping
can account for the observed extraplanar H$\alpha$ and H{\sc i} emission.
Oosterloo \& van Gorkom (2005) observed NGC~4388 with the WSRT and discovered an H{\sc i} tail which
extends more than $100$~kpc to the north.

We clearly detect X-ray emission associated to the H{\sc i} and H$\alpha$
tail of NGC~4388. We can only attribute the observed X-ray emission to the first $\sim 5'$ $(\sim 25$~kpc) of the H{\sc i} tail
which is about 5 times longer. This is because the projected H{\sc i} distribution approaches M~86 and the
halo of M~86 dominates the X-ray emission in this region. 

\begin{figure}[ht]
\resizebox{\hsize}{!}{\includegraphics{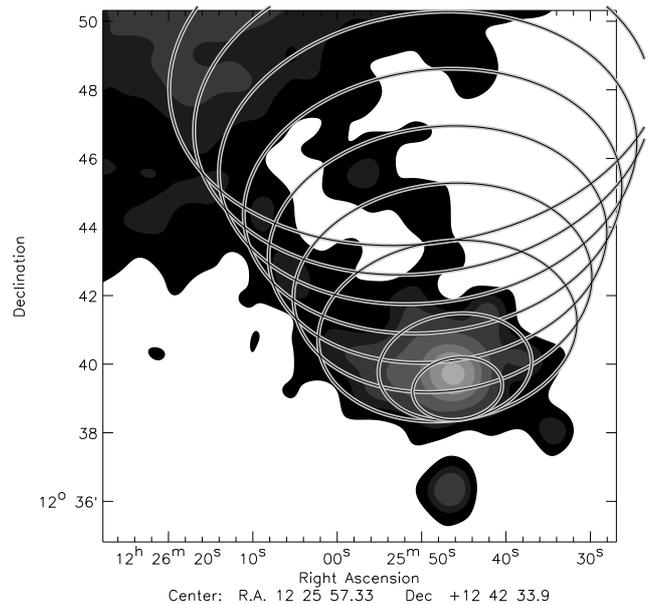}}
\caption{X-ray emission (0.2-1~keV) of NGC~4388 together with the Mach cone
whose direction is based on the dynamical model and whose opening angle is adjusted to the X-ray distribution.}
\label{fig:n4388m}
\end{figure}

As for NGC~4569 we adjusted a Mach cone to the X-ray and H{\sc i} emission distribution with the direction being
based on the dynamical model (Vollmer \& Huchtmeier 2003, Vollmer 2009). The apparent Mach cone angle is 
$\alpha \sim 26^{\circ}$ (Fig.~\ref{fig:n4388m}). With a normalized 3D model velocity vector (0.09,-0.75,0.66)
(Vollmer 2009) the true Mach cone angle is $20^{\circ}$.
This corresponds to a Mach number of $2.9$ and a galaxy velocity of $\sim 1600$-$2000$~km\,s$^{-1}$ (with a sound speed
of $550$-$700$~km\,s$^{-1}$), which is consistent with the estimated galaxy velocity based on the dynamical model 
($1700$~km\,s$^{-1}$, Vollmer 2009).
To obtain an Alfv\'{e}nic Mach number of $2.9$ with $v_{\rm gal}=1700$~km\,s$^{-1}$, 
a magnetic field strength of $\sim 3$~$\mu$G is needed with an intracluster density of $n=10^{-4}$~cm$^{-3}$. 

Within the X-ray tail we measure a temperature of $0.87$~keV ($10^7$~K) and a density of $n \sim \eta^{-0.5}\,10^{-3}$~cm$^{-3}$.
The thermal pressure is thus $p_{\rm th}=1.4 \times 10^{-12}$~dyn\,cm$^{-2}$, whereas ram pressure is about
$4$-$8 \times 10^{-12}$~dyn\,cm$^{-2}$. As in the halo of NGC~4569, the thermal pressure based on the X-ray observations is
significantly lower than the estimated ram pressure acting on NGC~4388.

We made an attempt to calculate the total mass of the hot X-ray gas within the whole H{\sc i} tail. 
Unfortunately, this is not directly possible due to the dominating X-ray halo of M~86.
Several assumptions have to be made: 
first, a constant surface-brightness along the whole tail is assumed. 
Next, the flux from the area of detectable X-ray tail is rescaled to the total volume of the H{\sc i} tail.
Since the H{\sc i} tail is $\sim 3$ times longer than the X-ray tail, this leads to a 3 times larger volume.

Given the temperature from the spectral fit, the density of the hot gas is $n \sim \eta^{-0.5} 10^{-3}$~cm$^{-3}$.
%The total mass of the hot gas associated to the X-ray emitting region is thus $\leq 3\,10^8$~M$_{\odot}$.
A total gas mass of $\sim 2 \times 10^8$~M$_{\odot}$ is found in the part of the X-ray tail which is not
dominated by M~86's halo (south to the declination $12\ 48\ 00$).
Since our volume estimate is on the high side and since we expect a volume filling factor smaller than 1, a more
realistic mass estimate of the X-ray tail is $\sim 10^8$~M$_{\odot}$. 
Within a 3 times longer tail, there might be a few $10^{8}$~M$_{\odot}$
of hot gas associated with the entire H{\sc i} tail of NGC~4388.
This total mass of hot gas is comparable to the total mass of atomic hydrogen in the tail
($3.4 \times 10^{8}$~M$_{\odot}$; Oosterloo \& van Gorkom 2005). The total gas mass of the tail 
($\sim 6 \times 10^{8}$~M$_{\odot}$) is much lower than the expected stripped gas mass based on the H{\sc i} deficiency
(2$\times$10$^9$M$_{\sun}$; Vollmer \& Huchtmeier 2007).
 
Possible reasons for this discrepancy are strong ISM-ICM mixing with a strong decrease of the gas density or
star formation within the gas tail. 
Based on the reconstructed star formation history from deep optical spectra Pappalardo et al. (2010) 
found that the stripped gas left the galactic disk $\sim 200$~Myr ago. The gas which has been stripped from
the location where the optical spectra were taken is presumably now in middle of the H{\sc i} tail. 
The gas within the tail thus
traveled with a velocity in the sky plane of $\sim 300$~km\,s$^{-1}$. With a radial velocity of
$\sim 300$~km\,s$^{-1}$ the total outflow velocity of the stripped gas is $\sim 400$-$450$~km\,s$^{-1}$.
This 3D configuration is consistent with the Vollmer \& Huchtmeier (2003) model.
The stripping timescale is reasonable in a scenario where NGC~4388 has been stripped by a recent passage through
the Virgo cluster core, close to M~87 (Vollmer 2009). Therefore, we think that the tail traces the entire 
stripping process. Given the small number of H{\sc ii} regions within the H{\sc i} tail (Yoshida et al. 2004), gas loss
via star formation seems not to be significant.

Vollmer \& Huchtmeier (2007) suggested an alternative explanation for the missing stripped gas:
spiral galaxies entering a cluster may rapidly lose the atomic gas which 
is located beyond the optical radius. This gas is (i) less bound than the gas in the inner disk and (ii) no longer 
stirred by strong interstellar turbulence and can easily be heated and evaporated.
The stripping of the outer gas disk might happen before the galaxy enters the cluster core.
In this case the galaxy already shows an initial H{\sc i} deficiency of $0.4$ before the stripping event.
If such an initial H{\sc i} deficiency is assumed, the sum of the observed H{\sc i} and extrapolated X-ray
gas mass might account for the whole stripped gas. Alternatively, a fraction of the gas mass might be
in the form of tenuous ionized gas with temperatures below $10^6$~K.

\subsection{NGC~4501}

Polarized radio emission studies showed a strong compression of the magnetic field at the southwestern edge of the galactic disk 
(We\.zgowiec et al.~\cite{wezgowiec}, Vollmer et al.~\cite{letter}).
NGC~4501's H{\sc i} disk is sharply truncated to the southwest, well within the stellar disk. 
On the opposite side, there is low surface-density gas, which is more extended than the main H{\sc i} disk.
A detailed comparison of observations with a sample of dynamical simulations, showed that ram pressure stripping can account 
for the main observed characteristics of NGC~4501 (Vollmer et al. 2008). 
The galaxy is stripped nearly edge-on, is heading southwest, and is $200$-$300$~Myr before 
peak ram pressure, i.e. its closest approach to M~87.

\begin{figure}[ht]
\resizebox{\hsize}{!}{\includegraphics{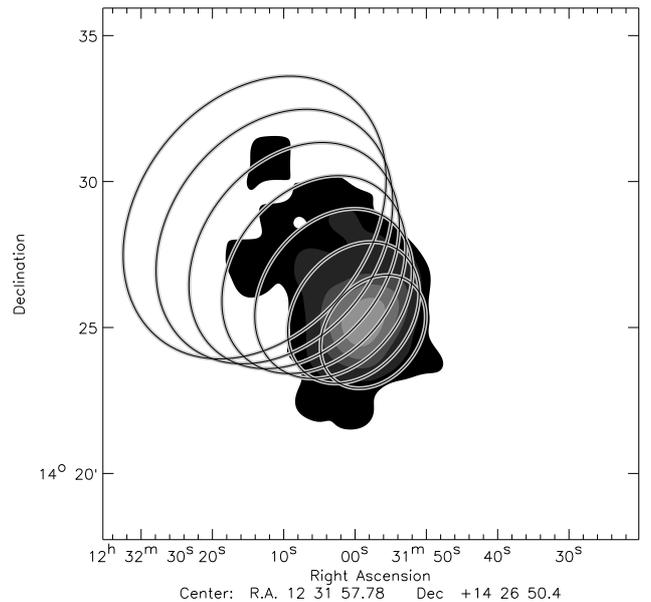}}
\caption{X-ray emission (0.2-1~keV) of NGC~4501 together with the Mach cone
whose direction is based on the dynamical model and whose opening angle is adjusted to the X-ray distribution.}
\label{fig:n4501m}
\end{figure}

A possible Mach cone is adjusted to the X-ray emission distribution in Fig.~\ref{fig:n4501m}.
The apparent Mach cone angle is $\sim 30^{\circ}$. 
With a normalized 3D model velocity vector (0.48,-0.44,0.76) (Vollmer 2009) the true Mach cone angle is $20^{\circ}$
implying a Mach number of $M \sim 2.9$. Assuming a lower angle between the galaxy's normalized 3D velocity vector and 
the plane of the sky ($\beta=37^{\circ}$ instead of $49^{\circ}$ yields a true Mach cone angle of $25^{\circ}$ and
a Mach number of $M \sim 2.4$. If the sound speed is
relevant for the Mach number the galaxy velocity is $v_{\rm gal} \sim 1300$-$2000$~km\,s$^{-1}$. This is consistent with
the galaxy velocity based on the dynamical model ($v_{\rm gal}=1500$~km\,s$^{-1}$).
If the Alfv\'{e}n velocity exceeds the sound speed, a magnetic field strength of $\sim 2$-$3$~$\mu$G is needed
assuming an intracluster medium density of $n=10^{-4}$~cm$^{-3}$.

Our X-ray data do not provide enough counts to obtain a high signal-to-noise spectrum. Therefore, we fitted a
single temperature model to the X-ray spectrum of the tail and found the temperature to be $\sim 0.7$~keV. 
This temperature is comparable to
the temperatures of the hot extraplanar gas in NGC~4569 ($0.49$~keV) and NGC~4388 ($0.87$~keV).
The X-ray emission beyond the H{\sc i} disk might be due to stripped atomic gas mixed with the intracluster medium or
a galactic wind due to star formation in the prominent northwestern spiral arm (Fig.~\ref{fig:n4501x}).

\subsection{Hot gas within the Mach cones}

For NGC~4569 we have the lucky situation that the Mach cone is entirely filled by a galactic outflow.
A galactic outflow encountering the ICM flow within the Mach cone should produce an observable
contact discontinuity (see, e.g., Heinz et al. 2003, Kraft et al. 2011) in addition to the bow shock.
According to Kraft et al. (2011) and Vikhlinin et al. (2001) the separation between the two features
is $\sim 0.15$ times the distance between the galaxy center and the contact discontinuity for large Mach numbers. 
Assuming this distance to be $25$~kpc, the projected distance between the bow shock and a contact discontinuity
is about $45''$. This is close to the image resolution of $1'$. The fact that we do not observe two
distinct features might be caused by the absence of a contact discontinuity, because the ICM flow
within the Mach cone is turbulent, or might be due to resolution/sensitivity and/or projection effects.
%and can thus be recognized by its triangular shape.

On the other hand, the Mach cone geometries of NGC~4388 and NGC~4501 are tentative.
All extraplanar X-ray emitting hot gas is found within the Mach cones of the three galaxies.

Whereas it is most likely that the diffuse X-ray halo of NGC~4569 (Fig.~\ref{fig:n4569x}) is due to a galactic superwind
originating in a nuclear starburst, the extraplanar X-ray emission of NGC~4388 is most likely due 
to the mixing of stripped ISM with the hot intracluster medium. 
Typical temperatures of the cold and hot phase in the star formation driven outflow of NGC~253 are 
$\sim 0.1$~keV and $\sim 0.3$~keV, respectively (e.g. Bauer et al. 2008).
The appearance of a ``multi-temperature'' halo is a natural consequence of the plasma driven out of collisional ionization 
equilibrium (CIE) towards a non-equilibrium ionisation (NEI) state in a galactic outflow.
The temperature of the hottest component in galactic outflows of massive spiral galaxies is about $0.4$~keV (Martin 1999).

The gas in NGC~4388's and NGC~4501's tails is significantly hotter than that of typical galactic outflows
of massive spiral galaxies.
On the other hand, the X-ray tail of ESO~137-001 in ACO~3627 has a temperature of $0.7$~keV (Sun et al. 2006).
The temperature of the X-ray tail of ESO~137-002 might be as high as $2$~keV (Sun et al. 2010). 
Both galaxies evolve in a galaxy cluster whose intracluster medium is 3 times hotter than that of the Virgo cluster
($2.4$~keV, B\"{o}ringer et al. 1994). 
We suggest that the higher gas temperatures observed in the X-ray tails of NGC~4388 ($0.9$~keV) and 
NGC~4501 ($0.7$~keV) are due to the mixing of the stripped ISM into the hot intracluster medium of the Virgo cluster.

\section{Conclusions \label{sec:conclusions}}

XMM-Newton observations of three Virgo cluster spirals are presented. These galaxies are part of the
ram pressure stripping time sequence of Vollmer (2009): NGC~4501 is still approaching the cluster center,
NGC~4388 and NGC~4569 have experienced peak ram pressure $\sim 150$~Myr ago and $\sim 300$~Myr ago.
We find extraplanar diffuse X-ray emission in all galaxies:
\begin{itemize}
\item 
In NGC~4569 it is associated with the galactic outflow observed in H$\alpha$ and radio continuum (Chy\.zy et al. 2006). 
The diffuse X-ray emission has a triangular shape.
\item
In NGC~4388 the X-ray emission follows the beginning of the H{\sc i} tail discovered by Oosterloo \& van Gorkom (2003). 
The halo of M~86 dominates the X-ray emission at larger distances from the galaxy. 
\item
In NGC~4501 it is located opposite to the side of the galactic disk where the gas is compressed by ram pressure
(Vollmer et al. 2008).
\end{itemize}
Based on the 3D velocity vectors from dynamical modelling (Vollmer 2009) a simple Mach cone can be
fitted to the triangular shape of NGC~4569's diffuse X-ray emission.
To our knowledge this is the first direct evidence for a Mach cone in a cluster spiral galaxy.
We can only observe it, because it is filled with hot gas from a galactic superwind. 
Assuming that all extraplanar diffuse X-ray emission has to be located inside the Mach cone, we also
fit Mach cones to NGC~4388 and NGC~4501. 

For NGC~4569 it is hard to reconcile the derived Mach cone
opening angle with a Mach number based on the sound speed alone. Instead, a Mach number
involving the Alfv\'enic speed seems to be more appropriate, yielding a magnetic field strength of $\sim 3$-$6$~$\mu$G
for a intracluster medium density of $n \sim 10^{-4}$~cm$^{-3}$.
The shock is thus caused by fast magnetosonic waves.
%It might thus be the magnetic field which makes the intracluster medium collisional as it is the case
%for the solar wind.

Gas densities and temperatures are derived from the X-ray spectra. 
In the outflow of NGC~4569 the thermal pressure is a factor of 2 smaller than the cosmic ray gas pressure.
The outflow might thus be driven by cosmic rays. 
The thermal pressure in NGC~4388's X-ray tail is significantly smaller than the ram pressure which NGC~4388 is experiencing.
Whereas the temperature of the hot
component of NGC~4569's X-ray halo ($0.5$~keV) is at the high end but typical for a galactic outflow, the temperature
of the hot gas tails of NGC~4388 and NGC~4501 are significantly hotter ($0.7$-$0.9$~keV). 
We suggest that these higher gas temperatures are due to the mixing of the stripped ISM into the hot intracluster 
medium of the Virgo cluster.

\begin{acknowledgements}
This work was supported by DLR Verbundforschung
"Extraterrestrische Physik" at Ruhr-University Bochum through
grant 50 OR 0801.
\end{acknowledgements}

\begin{appendix}

\section{X-ray emitting volumes}

\begin{figure}[ht]
\begin{center}
\resizebox{6cm}{!}{\includegraphics[angle=0]{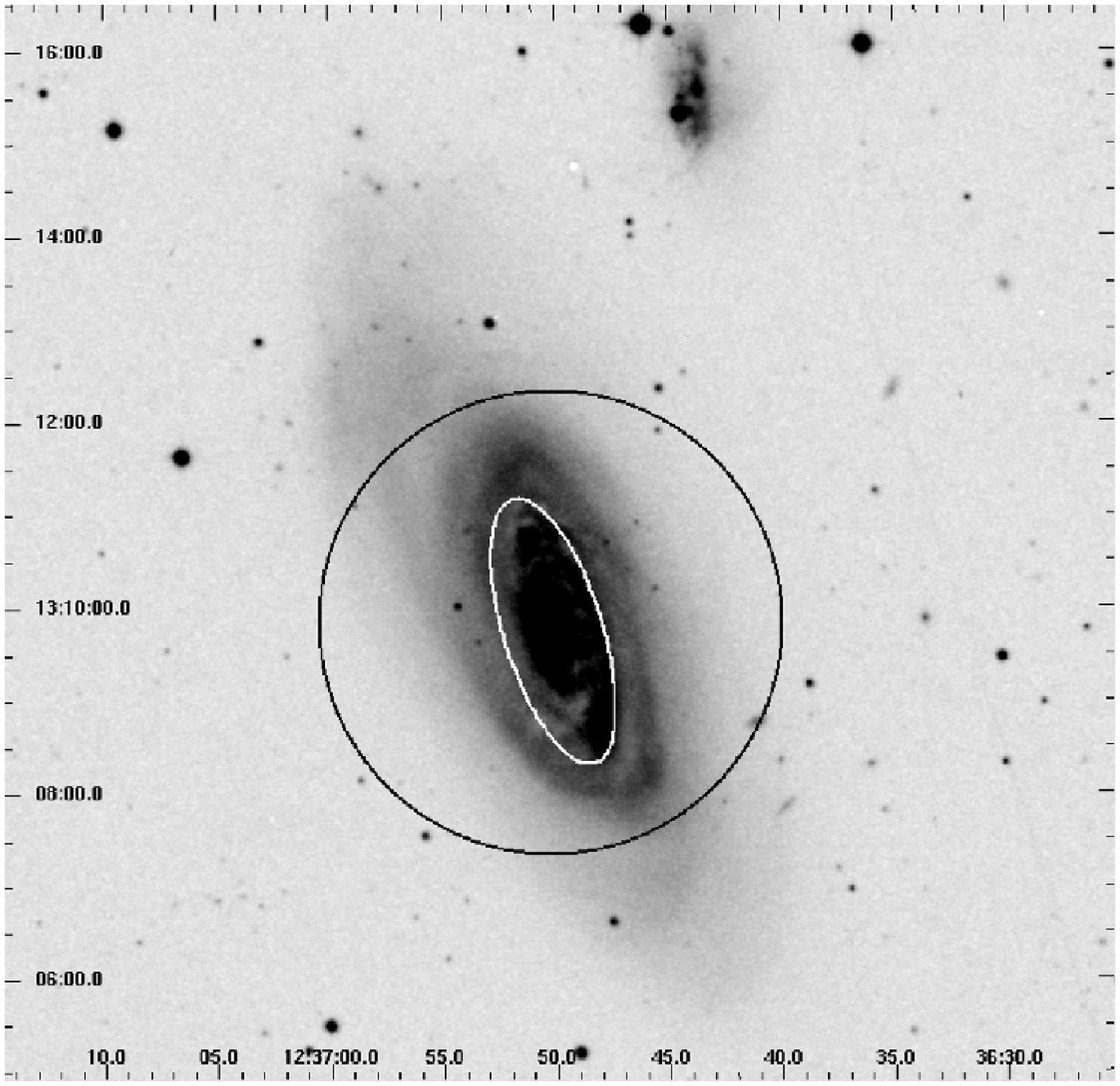}}
\resizebox{6cm}{!}{\includegraphics[angle=0]{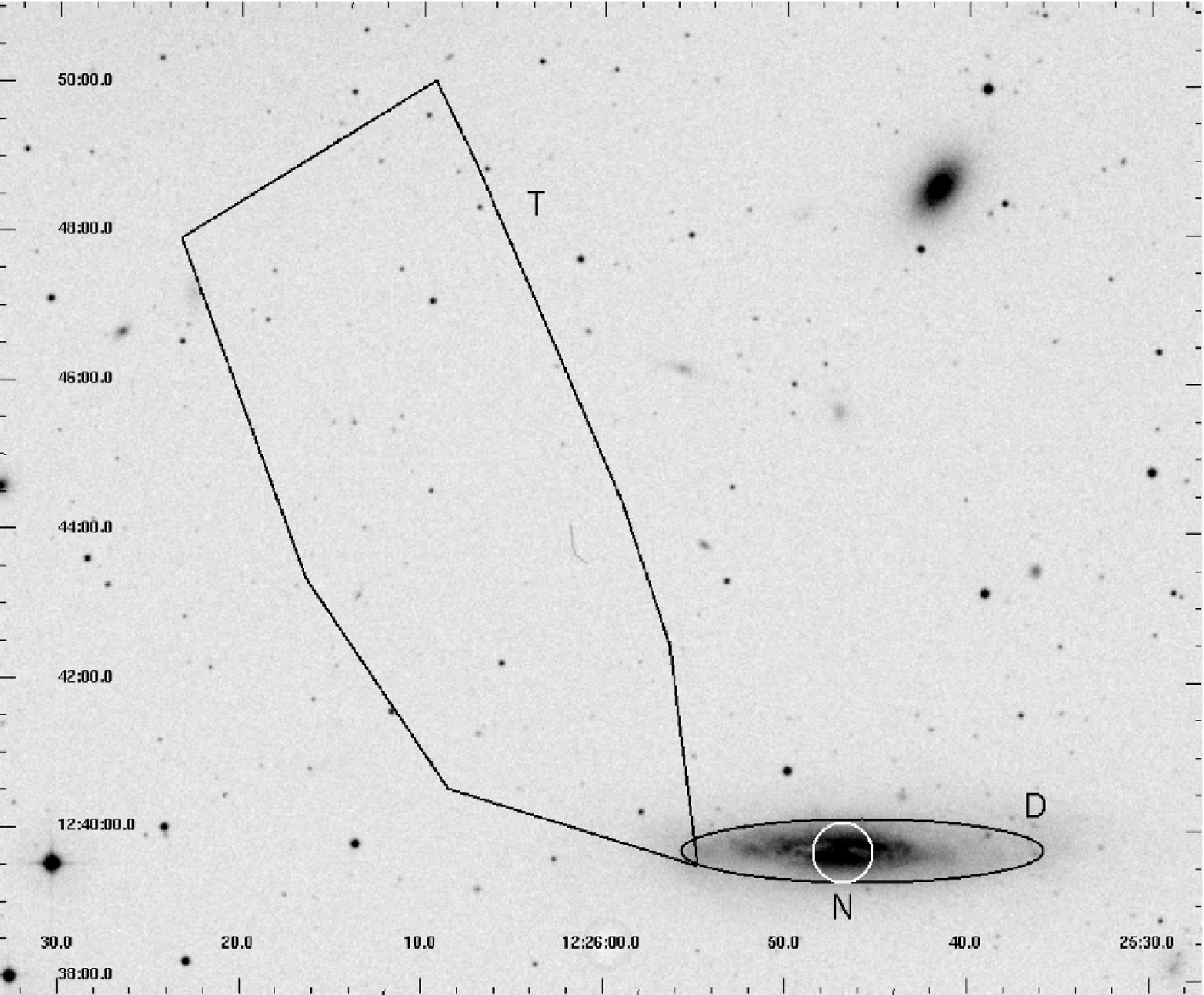}}
\resizebox{6cm}{!}{\includegraphics[angle=0]{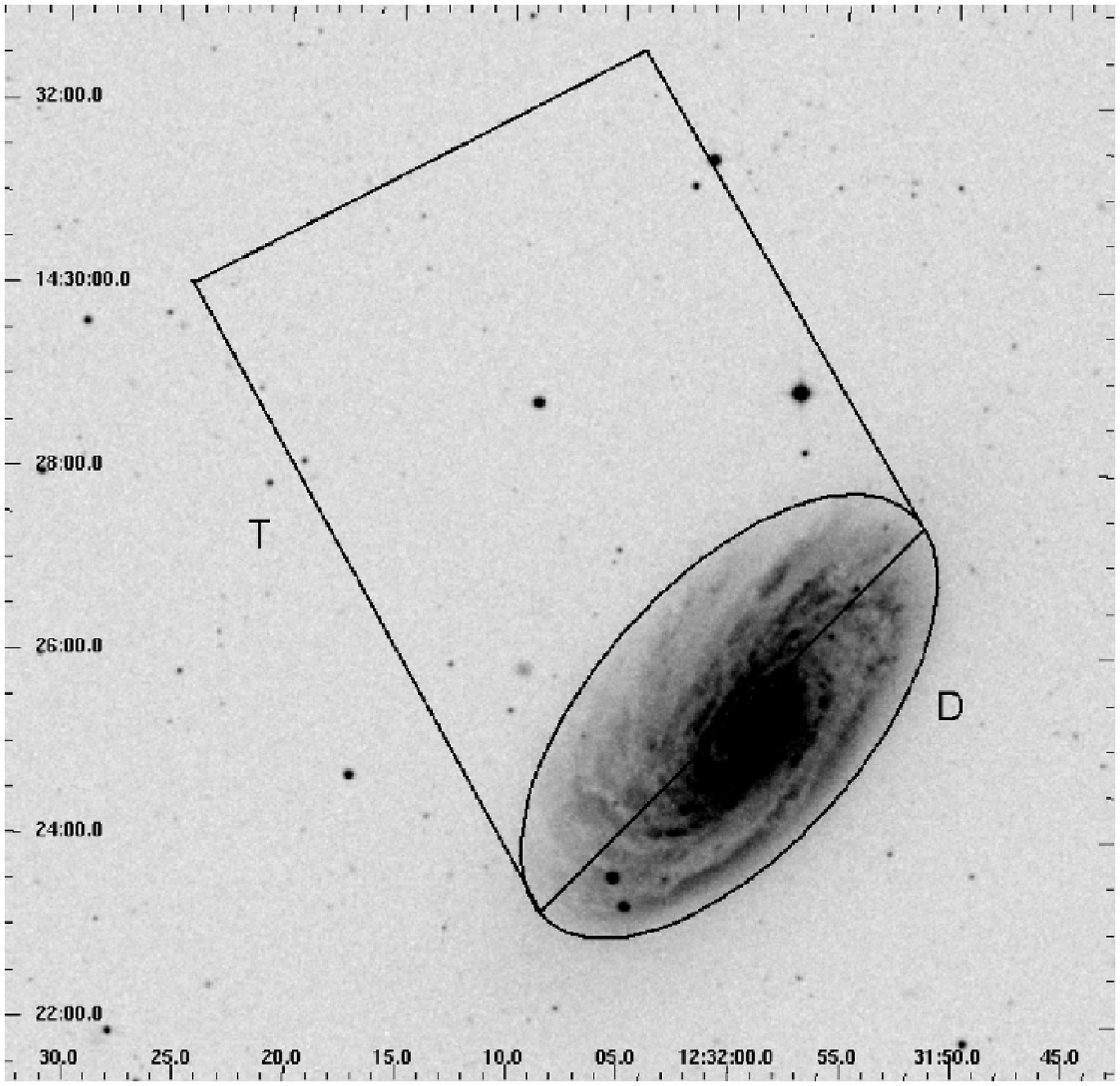}}
\end{center}
\caption{Regions for X-ray spectrum extraction and X-ray emitting volume calculations.}
\end{figure}

\end{appendix}


\begin{thebibliography}{}

\bibitem{a0} Arnaud, K.A., 1996, Astronomical Data Analysis Software and Systems V, 
eds. Jacoby G. and Barnes J., p17, ASP Conf. Series volume 101

\bibitem{a0a} Bauer, M., Pietsch, W., Trinchieri, G., et al. 2008, A\&A, 489, 1029

\bibitem{a0b} B\"{o}hringer, H., Briel, U.G., Schwarz, R.A., et al. 1994, Nature, 368, 828

\bibitem[2006]{boselli06} Boselli A., Boissier S., Cortese L., et al. 2006, \apj, 651, 811

\bibitem[2007]{carter} Carter, J., A., \& Read, A., M., 2007, A\&A, 464, 1155

\bibitem[1990]{cayatte} Cayatte, V., van Gorkom, J.~H., Balkowski, C., \& Kotanyi, C., 1990, AJ, 100, 604

\bibitem[2007]{chung} Chung, A., van Gorkom, J., H., Kenney, J., D., P., \& Vollmer, B., 2007, A\&A, 659, 115

\bibitem{a1} Chung, A., van Gorkom, J.H., Kenney, J.D.P., Crowl, H., \& Vollmer, B. 2009, AJ, 139, 2716

\bibitem[2006]{chyzy4569} Chy\.zy, K., T., Soida, M., Bomans, D., J., et al., 2006, A\&A, 447, 465

\bibitem{a2} Ferrari, C., Govoni, F., Schindler, S., Bykov, A.M., Rephaeli, Y. 2008, SSRv, 134, 93

\bibitem{a3} Finoguenov A., Briel, U.G., Henry, J.P., et al. 2004, A\&A, 419, 47

\bibitem[2004]{sas} Gabriel, C., Denby, M., Fyfe, D., J., et al., 2004, ASPC, 314, 759

\bibitem{a3a} Heinz, S., Churazov, E., Forman, W., Jones, C., \& Briel, U.G. 2003, MNRAS, 346, 13

\bibitem[1992]{kaastra} Kaastra, J. S. 1992, An X-Ray Spectral Code for Optically Thin Plasmas (Internal SRON-Leiden Report, updated version 2.0)

\bibitem[2005]{lab} Kalberla, P., M., W., Burton, W., B., Hartmann D., et al., A\&A, 440, 775

\bibitem{a3b} Kraft, R.P., Forman, W.R., Jones, C., et al. 2011, ApJ, 727, 41

\bibitem{a3c} Martin, C.L. 1999, ApJ, 513, 156

\bibitem[1985]{mewe} Mewe, R., Gronenschild, E. H. B. M., \& van den Oord, G. H. J. 1985, \aaps, 62, 197

\bibitem[2005]{oosterloo} Oosterloo, T., \& van Gorkom, J., A\&A, 437, L19

\bibitem[2010]{pappalardo} Pappalardo, C., Lan\c{c}on, A., Vollmer, B., et al. 2010, A\&A, 514, 33

\bibitem[2003]{leda} Paturel, G., Petit, C., Prugniel, P., et al., 2003, A\&A, 412, 45

\bibitem{a3d} Rasmussen, J., Ponman T.J., \& Mulchaey J.S. 2006, MNRAS, 370, 453 

\bibitem{a4} Roediger, E., Br\"{u}ggen, M., \& Hoeft, M. 2006, MNRAS, 371, 609 

\bibitem{a5} Roediger, E., \& Br\"{u}ggen, M. 2008, MNRAS, 388, 465

\bibitem{a6} Schulz, S., \& Struck, C. 2001, MNRAS, 328, 185
 
\bibitem[2006]{soida06} Soida M., Otmianowska-Mazur K., Chy{\.z}y K., Vollmer B., 2006, A\&A, 458, 727

\bibitem{a14} Solanes J.M., Manrique A.,  Garcia-Gomez C., et al. 2001, ApJ, 548, 97

\bibitem{a7} Stevens, I.R., Acreman, D.M., \& Ponman, T.J. 1999, MNRAS, 310, 663

\bibitem[2001]{strueder} Str\"uder, L., Briel, U., Dennerl., K., et al., 2001, A\&A, 365, 18

\bibitem{a8} Sun, M. \& Vikhlinin, A. 2005, ApJ, 621, 718

\bibitem{a9} Sun, M., Jones, C., Forman, W., et al. 2006, ApJ, 637, L81

\bibitem{a10} Sun, M., Donahue, M., R\"{o}diger, E., et al. 2010, ApJ, 708, 946

\bibitem{a10a} Tsch\"{o}ke, D., Bomans, D.J., Hensler, G., Junkes, N. 2001, A\&A, 380, 40

\bibitem{a11} Tonnesen, S. \& Bryan G.L. 2009, ApJ, 694, 789

\bibitem[2001]{turner} Turner, M., J., L., Abbey, A., Arnaud, M., et al., 2001, A\&A, 365, 27

\bibitem{a17} van den Bergh S. 1976, ApJ, 206, 883

\bibitem{a18} Veilleux, S., Bland-Hawthorn, J., Cecil, G., Tully, R.B.M., Scott T. 1999, ApJ, 520, 111

\bibitem{a18a} Vikhlinin, A., Markevitch, M., \& Murray, S.S. 2001, ApJ, 551, 160

\bibitem[2001]{vollmer01} Vollmer, B., Cayatte, V., Balkowski, C., \& Duschl, W., J., 2001, ApJ, 561, 708

\bibitem[2003]{vollmer03} Vollmer B., \& Huchtmeier W. 2003, \aap, 406, 427

\bibitem[2004]{vollmer04} Vollmer B., Balkowski C., Cayatte V., et al. 2004, \aap, 419, 35

\bibitem[2006]{vollmer06} Vollmer B., Soida M., Otmianowska-Mazur K., et al., 2006, A\&A, 453, 883 

\bibitem[2003]{vollmer07} Vollmer B., \& Huchtmeier W. 2007, A\&A, 462, 93

\bibitem[2008]{vollmer4501} Vollmer, B., Soida, M., Chung, A., et al., 2008, A\&A, 483, 89

\bibitem[2007]{letter} Vollmer, B., Soida, M., Beck, R., et al., 2007, A\&A, 464, L37

\bibitem[2009]{model} Vollmer, B., 2009, A\&A, 502, 427 

\bibitem{a12} Wang, Q.D., Owen, F., \& Ledlow, M. 2004, ApJ, 611, 821

\bibitem{a22} Warmels R.H. 1988, A\&AS, 72, 57 

\bibitem[2007]{wezgowiec} We\.zgowiec, M., Urbanik, M., Vollmer, B., et al., 2007, A\&A, 471, 93

\bibitem[2002]{yoshida2002} Yoshida, M., Yagi, M., Okamura, S., et al. 2002, ApJ, 567, 118

\bibitem[2004]{yoshida} Yoshida, M., Ohyama, Y., Iye, M., et al., 2004, AJ, 127, 90

\end{thebibliography}
\end{document}